# Simulating Nonlinearity in Quantum Neural Networks While Mitigating Barren Plateaus

Ding-Dang Yang

Department of Economics, National Chengchi University, Taipei, 116, Taiwan, R.O.C.

E-mail: 112258018@g.nccu.edu.tw

**Keywords**: Power Series Expansion, Parameterized Unitary Matrices, Quantum Convolutional Neural Networks (QCNNs), Nonlinearity Simulation, Singular Value Decomposition (SVD), Image Classification in Quantum Computing, Quantum Noise

## Abstract

Quantum neural networks (QNNs) encounter significant challenges in realizing nonlinear behavior and effectively optimizing parameters. This study addresses these issues by modeling nonlinearity through a Taylor series expansion, where the method uses tensor products to generate the series basis, and parameterized unitary matrices define the corresponding coefficients. This design substantially reduces quantum circuit depth compared to conventional methods that rely on parameterized quantum gates, thereby mitigating the barren plateau problem. A QNN was implemented and tested on the MNIST and Fashion MNIST datasets to evaluate the proposed method, achieving test accuracies of 98.7% and 88.3%, respectively. With noise added, the accuracy decreased slightly to 98.6% and 87.2%.

**Introduction**

Classical artificial neural networks (ANNs) have been widely applied in image recognition, natural language processing, and healthcare [1, 2]. These networks rely heavily on large-scale matrix operations and require nonlinear components such as activation functions and max pooling to model complex patterns [3-5]. Due to the need for massively parallel computations, the development of hardware accelerators such as GPUs and TPUs has significantly accelerated the advancement of ANN research.

Due to their inherent properties of superposition and entanglement, quantum computers naturally excel at matrix multiplication, which constitutes one of the fundamental operations in quantum computing. With $n$ qubits, a quantum computer can represent $2^n$ quantum states simultaneously, enabling the efficient computation of matrix operations with dimensions $(2^n \times 2^n) * 2^n$. Researchers can apply quantum computers to neural network architectures using parameterized quantum gates to adjust matrix elements. Quantum computers offer a significant advantage when performing extremely large matrix multiplications.

As a result, numerous studies on quantum neural networks have been published, focusing on applications in pattern recognition, optimization, and cryptography [6-14]. However, quantum neural networks still face several challenges. One key issue arises when making all matrix elements independently tunable, similar to classical neural networks or convolutional architectures. As the matrix dimensions increase, constructing such matrices requires quantum circuits of considerable depth. Excessive circuit depth makes parameter optimization extremely difficult, often preventing the loss function from reaching its minimum—a phenomenon known as the barren plateau problem [15-18]. Existing solutions—such as careful parameter initialization, layer-wise training, and shallow circuit design—can somewhat alleviate this issue, but often come at the cost of reduced expressiveness or scalability of the network [19, 20].

Another fundamental challenge of quantum circuits arises from their inherently unitary nature. Regardless of how the circuit is designed or whether ancillary qubits are included, the transformation between input and output states must follow unitary evolution. The output quantum state is always a linear combination of the input, making it impossible to implement nonlinear functions directly. Measurement is necessary to convert the output quantum state into classical information, and it can induce second-order nonlinear effects based on the input data. The probability distribution of the output quantum state can be obtained by performing a fixed number of shots.

A classical computer can then compute nonlinear function values based on these probabilities, which are fed back into the quantum computer as input. Repeating this process multiple times enables the simulation of more complex nonlinear functions. However, this process requires several times more shots than the original setup, increasing computational costs and resource consumption. Recent research has proposed quantum-compatible activation functions and polynomial expansions to address this challenge [21, 22]. While these techniques show promise, they often rely on deep quantum circuits and complex parameter tuning, increasing hardware demands and heightening the risk of error accumulation.

Tensor products are fundamental operations in quantum computing. This study uses tensor products to generate the terms of a Taylor series expansion, significantly reducing the number of quantum gates required and introducing nonlinear behavior. A unitary matrix is employed to generate the basis of the Taylor series expansion, helping to mitigate the barren plateau problem. Once optimized, these unitary matrices can be decomposed and implemented using quantum gates to realize the corresponding transformations within quantum circuits [23, 24].

To evaluate the proposed method's effectiveness, we implemented it within a Quantum Convolutional Neural Network (QCNN) and tested it on benchmark datasets, including MNIST and Fashion MNIST (FMNIST).

**Methods**

**Introducing Nonlinear Effects Using Orthonormal Bases.** Classical deep neural networks process input data using nonlinear activation functions. However, these functions cannot be directly represented by unitary matrices, making them unsuitable for quantum circuit implementation. Equations (1)–(3) illustrate how deep neural networks compute weighted sums across multiple layers of neurons, where each neuron receives activation values from the preceding layer:

$$h^{(0)} = X \tag{1}$$

$$h^{(l)} = \sigma\left(W^{(l)} h^{(l-1)} + b^{(l)}\right) \quad l = 1, 2, ..., s. \tag{2}$$

The final output of the network is given by:

$$f(X) = h^{(s)} \tag{3}$$

However, these nonlinear activation functions do not directly correspond to quantum operations. To address this limitation, we employ a Taylor series expansion to approximate neural network functions, as represented in Equation (4):

$$f(X) = a + \sum_{m_0} a_{m_0} x_{m_0} + \sum_{m_0, m_1} a_{m_0 m_1} x_{m_0} x_{m_1} + \sum_{m_0, m_1, m_2} a_{m_0 m_1 m_2} x_{m_0} x_{m_1} x_{m_2} + ... \tag{4}$$

While $f$ is nonlinear concerning the raw data, it becomes linear when expressed on a power series basis. By transforming raw data into this basis and inputting it into a quantum computer, the resulting quantum states exhibit nonlinear effects compared to the original data. Furthermore, quantum circuits naturally generate these higher-order terms through tensor products, effectively simulating nonlinear features without explicitly implementing nonlinear functions. Equation (5) illustrates how these tensor products naturally form a higher-dimensional feature space:

$$X^{\otimes k} = \text{span}\left\{x_{m_0} x_{m_1} \ldots x_{m_{k-1}}\right\} \tag{5}$$

**Mitigating the Barren Plateau Problem Using Parameterized Unitary Matrices.** We use parameterized unitary matrices to alleviate the barren plateau problem as an alternative to conventional parameterized quantum gates. Traditional parameterized quantum gates contain matrix elements composed of sine and cosine functions. When multiple such gates are applied within a quantum circuit, their tensor products lead to increasingly complex compositions of these trigonometric functions. Stacking multiple layers of these gates further compounds the number of function products influencing the output quantum state.

Since the loss function is derived from measurement probabilities, it inherently depends on these compounded function products. Each sine and cosine function contains multiple local minima and maxima. Through repeated multiplication, these oscillatory functions exacerbate gradient vanishing, making it exponentially more difficult to locate the global minimum. This phenomenon is a fundamental cause of barren plateaus in deep quantum neural networks.

In contrast, when parameterized unitary matrices are used, the output quantum state follows a Taylor series expansion, as described in Equation 4. The basis functions in this expansion are constants, while the parameters of the unitary matrices serve as expansion coefficients (e.g., $a$, $a_{m0}$, $a_{m0m1}$, etc.). The optimization landscape becomes significantly smoother, as these coefficients are optimized directly rather than through compositions of trigonometric functions. This reduces the susceptibility to barren plateaus compared to conventional Quantum Neural Networks (QNNs) and deep neural networks.

Furthermore, constructing parameterized unitary matrices still requires multiple layers of parameterized quantum gates. As prior research has shown, the variance of the loss function's gradient decays exponentially with circuit depth [15]. Thus, reducing quantum circuit depth by replacing individual quantum gates with parameterized unitary matrices can effectively mitigate the barren plateau problem.

**QCNN Architecture.** The QCNN was employed to process image data, using an 8 × 8 pixel image as an example. Classical image data were encoded into quantum states using amplitude encoding via an encoder module, converting the image into a quantum state represented by six qubits [25, 26]. The encoded quantum state was then processed through quantum convolutional operations using quantum filters (Qfilters). Each Qfilter was implemented as a parameterized 16 × 16 unitary matrix, functioning similarly to convolutional kernels in classical convolutional neural networks. To ensure that the Qfilters satisfied the unitary matrix requirement, we applied Singular Value Decomposition (SVD), decomposing the matrix into its U, Σ, and V* components. A valid unitary matrix was then constructed by multiplying the U and V* components [27].

The QCNN architecture supports configurations with one, two, or three quantum convolutional layers, enabling progressively deeper and more hierarchical feature extraction. After the convolutional layers, a measurement operation was performed to convert the quantum states into classical data. Figure 1 illustrates the QCNN architecture, highlighting the encoder, Qfilters, and kernel coverage over the input image. This visualization provides a clear overview of how quantum convolutional operations efficiently process image data within the quantum circuit.

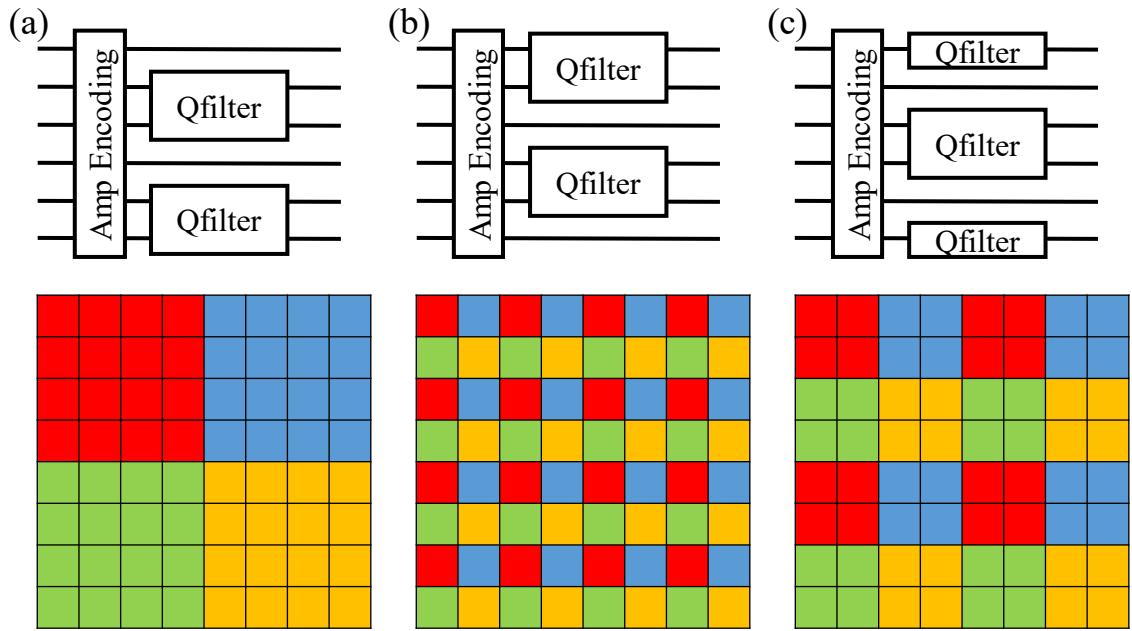

**Figure 1. QCNN Architecture.** This figure depicts the architecture of a QCNN designed to process an 8 × 8 pixel image using six qubits. The top panel illustrates the quantum circuit, where the encoder module uses amplitude encoding to convert classical image data into quantum states. Quantum convolutional operations are implemented using Qfilters, represented as 16 × 16 unitary matrices. Each Qfilter functions as a convolutional kernel with a size of 4 × 4, and 16 kernels are applied to extract features from the input data efficiently. The bottom panel visualizes the coverage of the convolutional kernels on the input image. The colored regions indicate distinct areas processed by each Qfilter, demonstrating various configurations for hierarchical feature extraction.

**Application of QCNN.** The QCNN was evaluated using the MNIST and FMNIST datasets, with image sizes resized to 8 × 8 pixels to meet input requirements. Quantum convolutional operations were implemented in PyTorch, leveraging its tensor operation capabilities for efficient matrix multiplications and parameter optimization. Following the quantum measurement step, the resulting classical data were processed through a Classical Fully Connected (CFC) layer with dimensions of 64 × 10. This layer mapped the outputs to one-hot encoded representations for classification tasks across ten categories.

Figure 2 illustrates the QCNN architectures used in this study, which include configurations with one, two, and three quantum convolutional layers. These hierarchical configurations allow for progressively deeper feature extraction, where additional layers capture increasingly complex patterns in the input data.

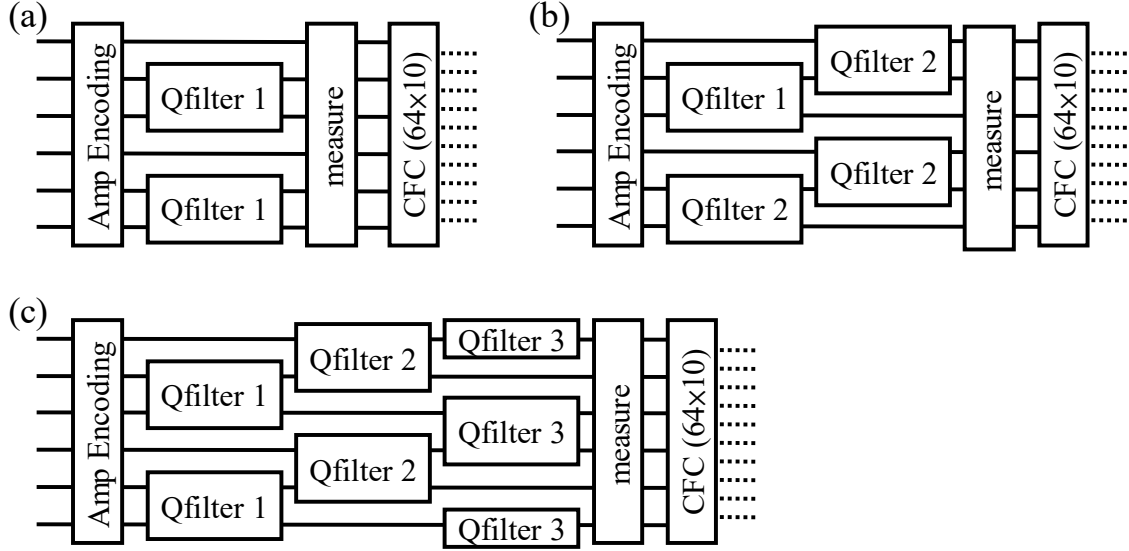

**Figure 2. QCNN with Increasing Convolutional Layers.** This figure illustrates the progression of QCNN architectures with one, two, and three quantum convolutional layers. In configuration (a), a single quantum convolutional layer is applied, followed by a measurement operation that converts quantum states into classical data. Configuration (b) adds a second quantum convolutional layer. Configuration (c) further enhances this process by incorporating a third quantum convolutional layer, capturing even more intricate patterns within the input data. In all configurations, the measurement operation outputs classical data, which is subsequently processed by a CFC layer with dimensions $64 \times 10$. The CFC layer uses one-hot encoding to map the outputs for classification across ten categories.

**Application of QCNN with Nonlinear Effects.** We introduce higher-order multiplicative terms of quantum states through tensor products to incorporate the nonlinear effects. This approach effectively simulates the role of activation functions in classical neural networks without explicitly implementing nonlinear functions. It leverages the intrinsic properties of quantum circuits while preserving the model's expressive power. The modified QCNN was evaluated on the MNIST and FMNIST datasets, with input images resized to $8 \times 8$ pixels to meet the network's dimensional requirements. Figure 3 illustrates the QCNN architecture with nonlinear effects, including configurations with one, two, and three quantum convolutional layers. In these configurations, tensor products generate pairwise multiplicative terms of the input quantum states, simulating the nonlinearity typically introduced by activation functions in classical neural networks. Input encoding uses amplitude encoding on two qubits, $q_{00}$ to $q_{05}$ and $q_{10}$ to $q_{15}$, enabling two independent encodings of the same MNIST or FMNIST image. These qubits collectively form a 12-qubit quantum circuit, ordered from highest to lowest significance as $q_{15}, q_{14}, q_{13}, q_{12}, q_{11}, q_{10}, q_{05}, q_{04}, q_{03}, q_{02}, q_{01}, q_{00}$.

Qubits $q_{15}$ through $q_{00}$ contain the binomial basis components of the Taylor series expansion (Equation 4). Qfilter 1 utilizes a subset of these basis elements — $q_{15}, q_{14}, q_{12}, q_{11}, q_{05}, q_{04}, q_{02}$, and $q_{00}$ — to capture higher-order interactions while maintaining computational efficiency. By selectively including only part of the basis, Qfilter 1 retains key polynomial terms essential for nonlinear feature extraction without incurring excessive circuit complexity.

As shown in configurations (b) and (c), deeper architectures incorporate additional Qfilter layers, progressively enhancing hierarchical feature extraction and representation learning. After quantum processing, a measurement operation collapses the quantum states into classical data. The measured outputs are subsequently passed to a CFC layer of dimension 4096 × 10, which maps the extracted quantum features into one-hot encoded labels for classification across ten categories.

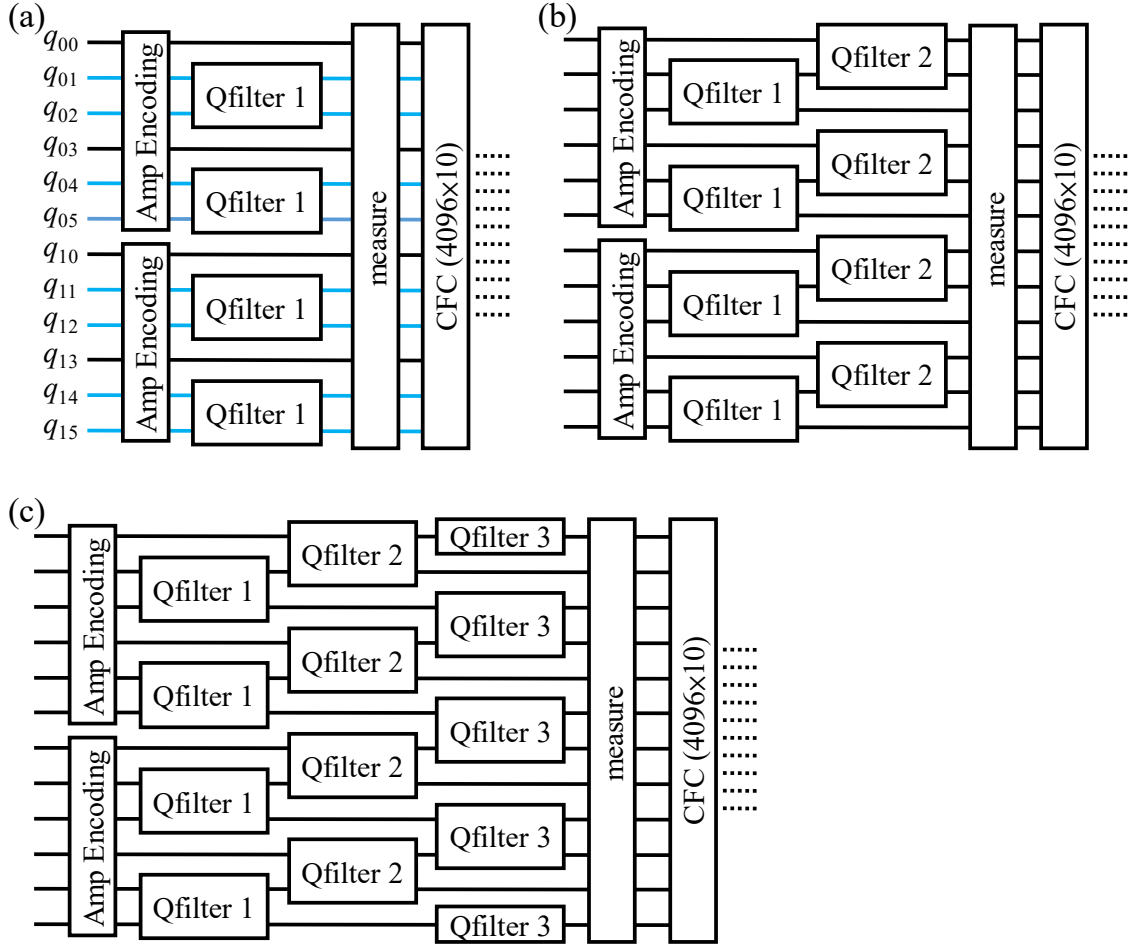

**Figure 3. QCNN Architecture with Nonlinear Effects.** This figure illustrates the QCNN architecture incorporating nonlinear effects through tensor products, which introduce multiplicative terms to simulate nonlinear activation functions. Three configurations are presented: (a) a single Qfilter layer, (b) two Qfilter layers, and (c) three Qfilter layers. In configuration (a), qubits $q_{00}$ to $q_{05}$ and $q_{10}$ to $q_{15}$ are used for amplitude encoding, forming a 12-qubit quantum circuit ordered from highest to lowest as $q_{15}$ to $q_{00}$. Qfilter 1 applies a linear combination of selected qubits to enhance feature extraction. Additional Qfilter layers in configurations (b) and (c) enable hierarchical feature learning, leading to progressively deeper quantum representations. After quantum processing, a measurement operation converts the quantum states into classical data, which is then processed by a CFC layer (4096 × 10) for classification.

**Implementation Details and Hyperparameter Settings.** Most hyperparameters were fixed to ensure model stability and reproducibility, while the learning rate was adjusted to optimize performance. The MNIST and FMNIST datasets—originally consisting of 28 × 28 grayscale images—were downsampled to 8 × 8 using bilinear interpolation. This reduction in image size helped lower computational complexity and meet the constraints of the quantum circuit. The pixel values were then normalized using L2 normalization to prepare the data for amplitude encoding, ensuring the input states satisfied the quantum unit norm constraint. The dataset was split into 60,000 training and 10,000 test images following standard practice.

A batch size of 100 was used for training. Stochastic gradient descent with a momentum of 0.9 was employed to accelerate convergence and reduce oscillations. The learning rate was empirically chosen to achieve the highest accuracy on the training set while maintaining model stability. The training was performed for 1,000 iterations, with accuracy recorded every 10 epochs for the training and test sets.

All computations—including preprocessing, training, and inference—were conducted on an Intel Core™ i7-14700 processor to ensure efficient execution. The resulting trained parameters were stored in a file for subsequent use. A separate program, built with PennyLane, then read these trained parameters to construct quantum circuits and verify that the results were consistent with those obtained during training.

To further assess the impact of quantum noise on accuracy, the trained parameters were used to incorporate depolarizing and phase-damping channels on all qubits, simulating realistic quantum noise. Noise levels were set to 0.05 for depolarizing and 0.03 for phase-damping on each qubit, capturing standard decoherence mechanisms typically encountered in physical quantum computers.

The dataset preprocessing pipeline, training logs, and evaluation scripts have been publicly available in a GitHub repository, allowing independent verification of the reported results.

**Results**

**Performance of Linear QCNNs.** During training, various learning rates were tested on the training dataset. The learning rate that achieved the highest training accuracy was selected for evaluating test performance (Figure A.1 in Appendix I for MNIST and Figure A.2 in Appendix I for FMNIST). Table 1 summarizes the performance of QCNNs with linear Qfilters. For the MNIST dataset, test accuracy steadily increased from 91.7% with one Qfilter layer to 94.1% with three layers. It demonstrates consistent performance gains as the network depth increases. Similarly, for the FMNIST dataset, test accuracy improved from 78.8% to 81.5% across three Qfilter layers. These findings suggest that while deeper linear QCNNs enhance data processing capabilities, the marginal performance gains diminish with additional layers.

**Performance of Nonlinear QCNNs.** As with the linear models, various learning rates were evaluated on the training dataset, and the one yielding the highest training accuracy was selected for testing (Figure A.3 in Appendix I for MNIST and Figure A.4 in Appendix I for FMNIST). Nonlinear QCNNs significantly outperformed their linear counterparts, as summarized in Table 1. For the MNIST dataset, test accuracy increased from 98.2% with one nonlinear Qfilter layer to 98.7% with three layers. Demonstrate a consistent performance improvement with increased depth. Similarly, for the FMNIST dataset, test accuracy improved from 87.1% to 88.3% as additional Qfilter layers were incorporated. These results highlight the effectiveness of nonlinear QCNNs in modeling complex data structures, leveraging higher-order terms derived from quantum operations to capture nonlinear effects.

**Impact of Quantum Noise on Accuracy.** To evaluate the effects of quantum noise on QCNN performance, we incorporated depolarizing (0.05) and phase-damping (0.03) noise channels into the quantum circuit. The results confirm that noise reduces test accuracy across all configurations, with linear QCNNs experiencing a more significant performance drop than nonlinear QCNNs (Table 1).

For the MNIST dataset, test accuracy in linear QCNNs decreased by approximately 1.4% to 2.0%. In contrast, nonlinear QCNNs exhibited a minor reduction of 0.1% to 0.6%, demonstrating superior noise resilience for the FMNIST dataset, where accuracy degradation was more significant in both linear and nonlinear QCNNs, suggesting that FMNIST is more sensitive to quantum noise than MNIST.

**Table 1. Performance Comparison of QCNNs with Linear and Nonlinear Qfilters on MNIST and FMNIST.** This table presents the training, standard test, and noise-affected test accuracies for QCNNs with 1 to 3 Qfilter layers on the MNIST and FMNIST datasets. Accuracy is reported in the format Train / Test / Noise-Test. The noise-affected accuracy (Noise-Test) was obtained by incorporating depolarizing (0.05) and phase-damping (0.03) noise channels into the quantum circuit to simulate realistic quantum hardware conditions.

| Qubits | Qfilter Layers | MNIST accuracy (Train/Test/Noise-Test) | FMNIST accuracy (Train/Test/Noise-Test) |
|---|---|---|---|
| 6 (linear) | 1 | 91.1 / 91.7/89.7 | 79.9 / 78.8/68.0 |
| | 2 | 93.2 / 93.7/91.9 | 81.8 / 80.9/78.9 |
| | 3 | 93.7 / 94.1/92.7 | 82.3 / 81.5/79.4 |
| 12 (nonlinear) | 1 | 99.0 / 98.2/ 97.6 | 90.7 / 87.1/ 82.9 |
| | 2 | 99.5 / 98.6/ 98.1 | 92.0 / 88.3/ 86.5 |
| | 3 | **99.5 / 98.7/ 98.6** | **92.3 / 88.3/ 87.2** |

**Comparison with Direct Pixel-to-CFC Mapping.** Table 2 summarizes the performance of models in which image features are directly mapped to a CFC layer without involving quantum processing. For the MNIST dataset, the test accuracy increased from 78.4% with an input size of 64 to 89.9% with 4096 input features. Similarly, for the FMNIST dataset, the test accuracy improved from 66.5% to 75.5%. These results suggest that increasing input dimensionality through direct pixel expansion can enhance model performance, although at the cost of significantly higher input size.

In contrast, nonlinear QCNNs achieved significantly higher accuracies. For the MNIST dataset, a single nonlinear Qfilter layer achieved 98.2% testing accuracy, surpassing the 89.9% obtained with 4096 input features in the classical mapping. Testing accuracy further increased to 98.7% with three nonlinear Qfilter layers. For the FMNIST dataset, a single nonlinear Qfilter layer achieved 87.1% testing accuracy, exceeding the 75.5% achieved with 4096 input features in the classical mapping, and improved further to 88.3% with three Qfilter layers.

These findings underscore the superior efficiency and representational power of nonlinear QCNNs in handling complex data structures. Unlike classical mappings, nonlinear QCNNs can achieve higher performance with fewer computational resources, making them a more effective solution for complex datasets.

**Table 2. Performance of Direct Pixel-to-CFC Mappings on MNIST and FMNIST.** This table summarizes the training and testing accuracies for direct mappings from image pixels to a CFC layer without involving quantum processing. Two input configurations are evaluated: 64 pixels and the tensor product of two 64-pixel vectors, resulting in a 4096-dimensional input. Results are reported as percentages in the format Train / Test for both MNIST and FMNIST datasets. The table highlights notable performance improvements when using the tensor product inputs, particularly with the 4096-dimensional representation, which achieves higher accuracy across both datasets.

| Input Size ($2^n$) | MNIST accuracy (Train/Test) | FMNIST accuracy (Train/Test) |
|---|---|---|
| $2^6 = 64$ | 77.5 / 78.4 | 67.2 / 66.5 |
| $2^{12} = 4096$ | 89.4 / 89.9 | 76.8 / 75.5 |

**Verification Using a Quantum Simulator.** We use PyTorch to optimize unitary matrix parameters efficiently while minimizing computational costs. This framework allowed for iterative parameter adjustments in a classical computing environment, leveraging its computational efficiency and scalability for rapid prototyping. The results were validated using PennyLane, a software platform for simulating quantum circuits, to verify the consistency of the optimized parameters with quantum hardware. The validation confirmed that PyTorch's optimized parameters produced consistent quantum circuit behaviors when simulated in a quantum environment.

**Discussion**

**Addressing Nonlinearity and Barren Plateaus in Quantum Neural Networks with Tensor Product-Based Taylor Series Expansion and Parameterized Unitary Matrices.** This study proposes a tensor product-based Taylor series expansion approach to model the nonlinear effects commonly utilized in classical neural networks. By leveraging tensor products—a fundamental operation in quantum computing—the data basis can be expanded to effectively replace classical activation functions. The Qfilter quantum circuit is designed to generate the Taylor series expansion basis coefficients, as defined in Equation (4), while simultaneously capturing their interdependencies within the Qfilter layer. In contrast, using parameterized quantum gates typically results in deep quantum circuits. Moreover, the associated loss functions often contain complex compositions of sine and cosine products, which introduce numerous local minima. As the loss landscape is defined over quantum state probabilities, such complexity significantly increases the likelihood of encountering Barren Plateaus—regions where gradients vanish, rendering optimization ineffective. To address this issue, we replace parameterized quantum gates with parameterized unitary matrices. This architectural modification reshapes the output quantum state so that the Taylor series expansion coefficients (as described in Equation (4)) depend on only a few parameters. As a result, the complexity of the loss function is reduced, thereby lowering the risk of Barren Plateaus and improving the trainability of the model.

**Importance of Nonlinear Effects in QCNNs.** The results shown in Table 1 demonstrate the importance of nonlinear effects on improving QCNN performance. Nonlinear QFilters, incorporating second-order multiplicative terms, effectively emulate classical activation functions and enable more expressive modeling of complex data structures. This advantage is evident in the MNIST and FMNIST datasets, where nonlinear QCNNs achieve higher accuracy than linear QCNNs and classical pixel-to-CFC. These results highlight the need for nonlinear mechanisms to unlock the full potential of quantum neural networks on structurally complex data.

**Efficiency of Quantum Convolutional Layers.** Table 2 presents a comparative analysis of classification accuracies between classical pixel-to-CFC mappings and QCNNs on the MNIST and FMNIST datasets, considering two different input sizes: $2^6$ and $2^{12}$. For the $2^6$ input size, pixel-to-CFC mappings achieved 78.4% accuracy on MNIST and 66.5% on FMNIST. In contrast, QCNNs attained higher accuracies of 94.1% and 81.5% on MNIST and FMNIST, respectively. When the input size increased to $2^{12}$, pixel-to-CFC mappings improved to 89.9% for MNIST and 75.5% for FMNIST. However, QCNNs demonstrated superior performance with accuracies of 98.7% on MNIST and 88.3% on FMNIST. These results underscore quantum convolutional layers' enhanced efficiency and scalability in processing complex data structures, outperforming classical methods even with lower-resolution inputs.

**Taylor Series Expansion and Tensor Basis Representation.** Equation (4) defines the Taylor series basis, which includes first-order, second-order, and higher-order terms, capturing nonlinear interactions among the input variables. In Figure 2, only first-order terms are considered, whereas Figure 3 extends to second-order terms for enhanced representation. Let $X$ represent the original input data (e.g., pixel values) that vary across images. An additional constant, $a$, is introduced, which remains unchanged across all photos. Consequently, the input data is represented as $\{a, X\}$. The tensor basis $\{a, X\}^{\otimes m}$ generates elements containing first-, second, and higher-order terms. This formulation enables the model to capture more complex dependencies within the data, improving its expressiveness and representation capabilities.

**Quantum Convolutional Kernels.** The primary function of Qfilters is to perform quantum transformations on input data by generating the coefficients ($a$, $a_{m0}$, $a_{m0m1}$, …) in Equation (4), which correspond to the Taylor series expansion. When a convolutional layer operates on $m$ qubits, the associated unitary matrix has dimensions $2^m * 2^m$, forming a quantum convolutional kernel that processes $2^m$ data points and contains $2^m$ kernels.

The number of independent parameters in a real-valued unitary matrix of this form is given by $(2^m(2^m - 1))/2$. As the matrix size increases, so does the number of parameters, enhancing expressiveness and increasing computational cost. Larger Qfilters incorporate more qubits, expanding parameter capacity while introducing more significant computational overhead, particularly during singular value decomposition (SVD), which is required to extract unitary matrices. Moreover, decomposing trained unitary matrices into quantum gates further amplifies the number of necessary quantum operations.

**Robustness of QCNNs Under Noise: Insights from Real Quantum Hardware Implementation.** Implementing tensor product-based Taylor series expansions on real quantum hardware presents several practical challenges, including noise, decoherence, gate errors, and qubit connectivity limitations. As shown in Table 1, noise significantly affects quantum circuit performance. Linear QCNNs suffer more significant degradation of accuracy under noise conditions—for example, test accuracy for MNIST drops from 91.7% to 89.7% and FMNIST from 78.8% to 68.0% with a single QFilter layer. In contrast, nonlinear QCNNs, which use tensor product-based expansions across 12 qubits, demonstrate stronger resilience to noise. For instance, test accuracy for MNIST decreases only slightly from 98.7% to 98.6%, and FMNIST from 88.3% to 87.2%, even under noise-affected conditions. Moreover, increasing the number of Qfilter layers further mitigates the impact of noise. As depth increases from 1 to 3 layers, the drop in accuracy under noise becomes less pronounced. These results suggest that nonlinear QCNNs provide higher baseline accuracy, and their architecture, especially with more QFilter layers, is inherently more robust against the detrimental effects of quantum noise.

**Comparison with Classical Deep Learning Models.** Classical deep learning models, particularly CNNs, have demonstrated state-of-the-art performance in image classification tasks such as MNIST and FMNIST. CNNs achieve efficient hierarchical feature extraction through convolutional layers and leverage explicit nonlinear activation functions (e.g., ReLU, Sigmoid) to enhance representational capacity. Due to well-optimized architectures and GPU acceleration, CNNs can achieve high accuracy with relatively low computational overhead.

In contrast, our approach introduces nonlinearity through tensor product-based quantum transformations, eliminating the need for explicit activation functions. Theoretically, any nonlinear function can be approximated using a Taylor series expansion, as described in Equations (1-4). Within the quantum framework, we encode Taylor series coefficients directly through quantum convolutional layers, enabling an alternative approach to nonlinearity well-suited for quantum circuits. Unlike classical deep neural networks, which typically require deeper architectures to extract complex features, our method leverages quantum tensor products to construct feature representations in a shallower network, reducing susceptibility to the barren plateau problem—a key challenge in deep quantum networks.

While classical neural networks could approximate activation functions using Taylor series expansions instead of explicit nonlinear functions, doing so necessitates an exponentially increasing amount of input data, making it computationally infeasible. Unlike traditional methods, tensor products—fundamental to quantum computing—facilitate efficient construction and manipulation of Taylor series basis functions. This inherent capability allows quantum circuits to bypass the exponential data requirements faced by classical computers.

However, quantum computing lacks a direct mechanism for implementing activation functions analogous to those in classical neural networks. Consequently, our approach capitalizes on the inherent strengths of quantum hardware by employing tensor product expansions as a natural and scalable method for introducing nonlinearity into quantum neural networks.

**Incorporating Nonlinearity into QNNs.** QNNs face an inherent challenge due to the intrinsic linearity of quantum circuits, where quantum states always form linear superpositions of input states. Recent research has proposed various strategies to introduce nonlinearity, each with distinct advantages and limitations.

A common approach involves intermediate measurements, wherein classical hardware processes the measurement results before feeding them back into the quantum system. While this method introduces the desired nonlinearity, it significantly increases computational overhead due to frequent measurements and data reloading, leading to inefficiencies and additional noise accumulation.

Another widely used technique employs parameterized quantum gates to induce nonlinearity, where the output quantum state exhibits a nonlinear dependence on tunable parameters. While this approach eliminates the need for repeated measurements, it suffers from insufficient parameter entanglement, limiting its ability to capture complex nonlinear relationships effectively.

Classical preprocessing is another popular solution, leveraging CPUs or GPUs to extract nonlinear features before encoding them into quantum states. This method efficiently captures nonlinear relationships without heavily burdening quantum resources. However, the quantum advantage remains contingent on the ability of the quantum circuit to utilize the preprocessed information fully.

We propose an alternative approach that leverages tensor products to construct Taylor series expansion bases. This enables a generalizable representation of nonlinear functions without requiring explicit nonlinear encoding in the quantum circuit. This significantly reduces the need for additional quantum gates, thereby maintaining a shallower quantum circuit depth. More importantly, the method is conceptually and technically simple, avoiding the overhead of intermediate measurements while preserving richer parameter dependencies than parameterized gate-based methods.

However, a key trade-off is that this method requires multiple copies of the original qubits, effectively increasing the number of qubits used. While the number of quantum gates scales accordingly, the depth of the quantum circuit remains unchanged, mitigating potential issues related to barren plateaus and excessive circuit complexity. This feature makes our approach particularly suitable for near-term quantum hardware, where circuit depth is a limiting factor, but qubit availability may be relatively more flexible.

**Conclusions**

This study uses tensor products to generate Taylor series bases and parameterized unitary matrices to determine the coefficients of each basis function. This method significantly reduces circuit depth while enhancing expressivity by incorporating nonlinear transformations directly within quantum circuits. To test on the MNIST and FMNIST datasets, the proposed model achieved test accuracies of 98.7% and 88.3%, respectively. When noise was introduced, quantum circuits incorporating nonlinear functions exhibited greater resilience, underscoring the robustness of the approach.

**Competing interests**

The authors declare no competing interests.

**Data and Software Availability**

The data supporting the findings of this study are openly available on GitHub at the following URL: https://github.com/Ding-Dang-Yang-ROC/QCNNS.

Appendix I.

**Dependence of Learning Rate on Accuracy for Training and Test Datasets.** QCNNs were applied to the MNIST dataset (Figure A.1) and the FMNIST dataset (Figure A.2) using the quantum circuit shown in Figure 2. Similarly, using the quantum circuit depicted in Figure 3, QCNNs were applied to the MNIST dataset (Figure A.3) and the FMNIST dataset (Figure A.4). The learning rate was varied to optimize accuracy on the training dataset, and the corresponding trained parameters were subsequently evaluated on the test dataset.

The accuracies reported in Table 1 correspond to the parameter settings that yielded the highest training accuracy, with the associated test accuracies presented accordingly.

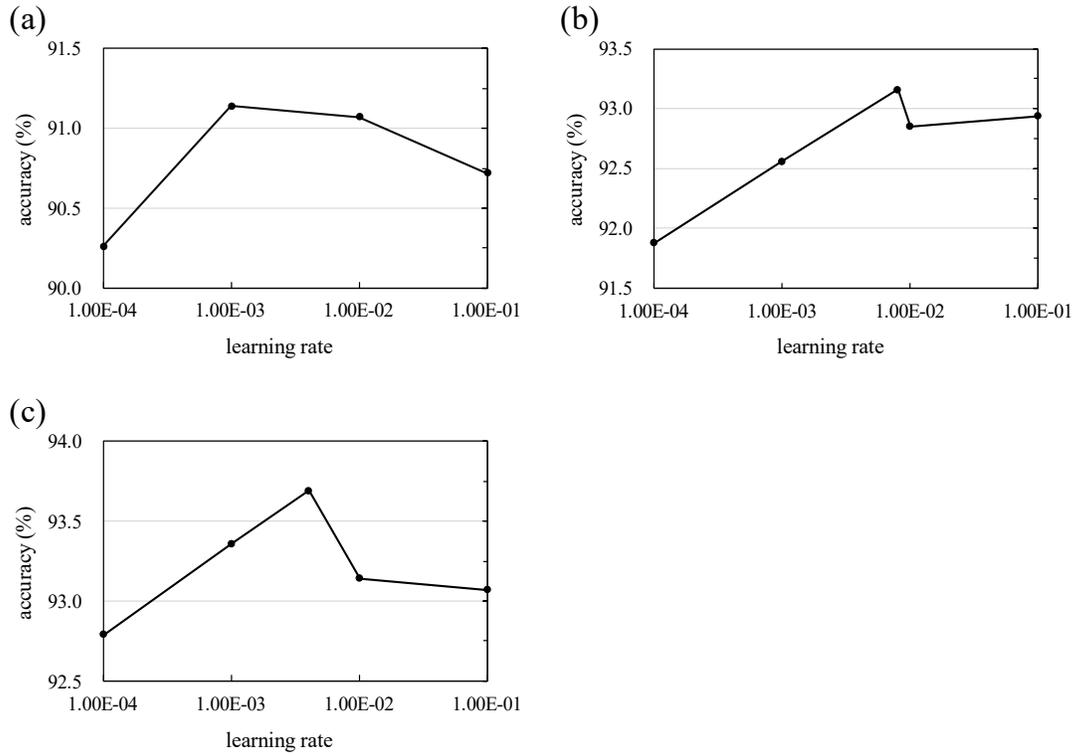

Figure A.1. QCNNs with a linear architecture (6 qubits) using the quantum circuit from Figure 2, evaluated on the MNIST dataset with (a) 1, (b) 2, and (c) 3 QFilter layers. The x-axis represents the learning rate, while the y-axis denotes accuracy.

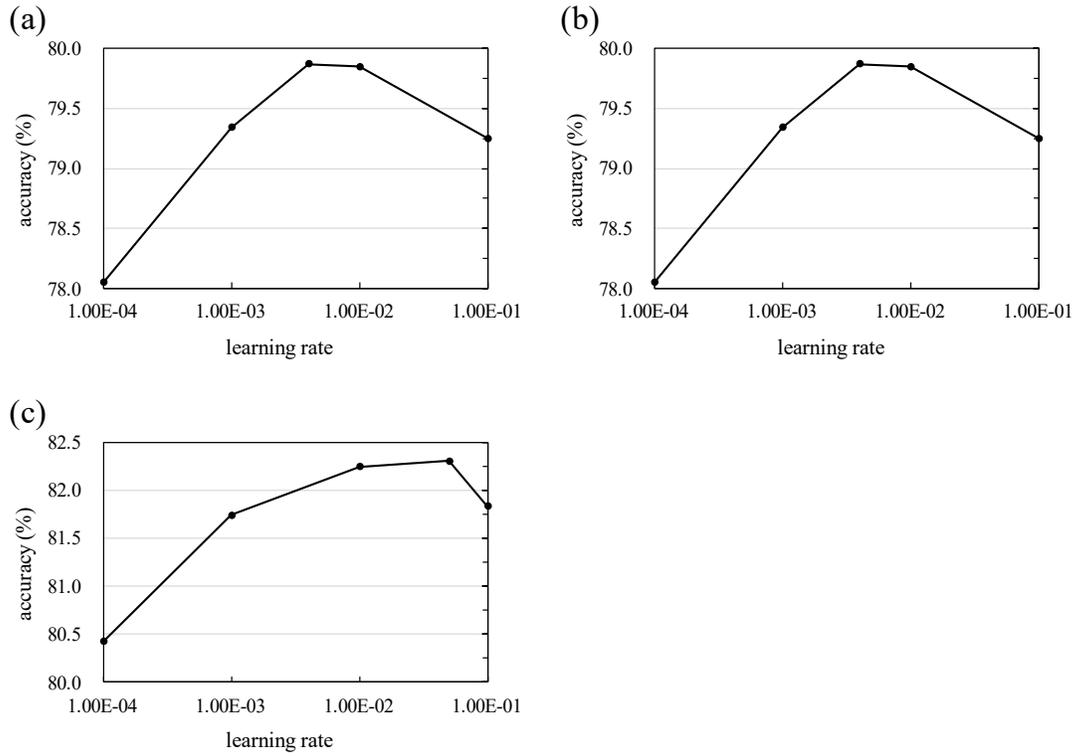

Figure A.2. QCNNs with a linear architecture (6 qubits) using the quantum circuit from Figure 2, evaluated on the FMNIST dataset with (a) 1, (b) 2, and (c) 3 QFilter layers. The x-axis represents the learning rate, while the y-axis denotes accuracy.

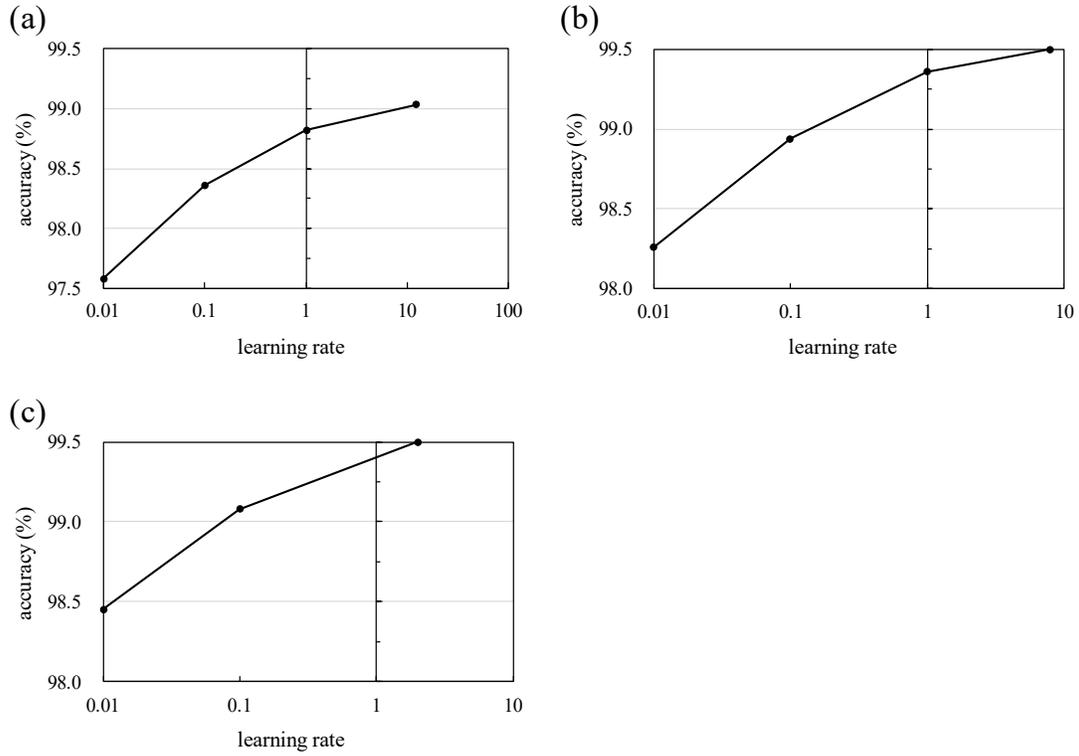

Figure A.3. QCNNs with a nonlinear architecture (12 qubits) using the quantum circuit from Figure 3, evaluated on the MNIST dataset with (a) 1, (b) 2, and (c) 3 QFilter layers. The x-axis represents the learning rate, while the y-axis denotes accuracy.

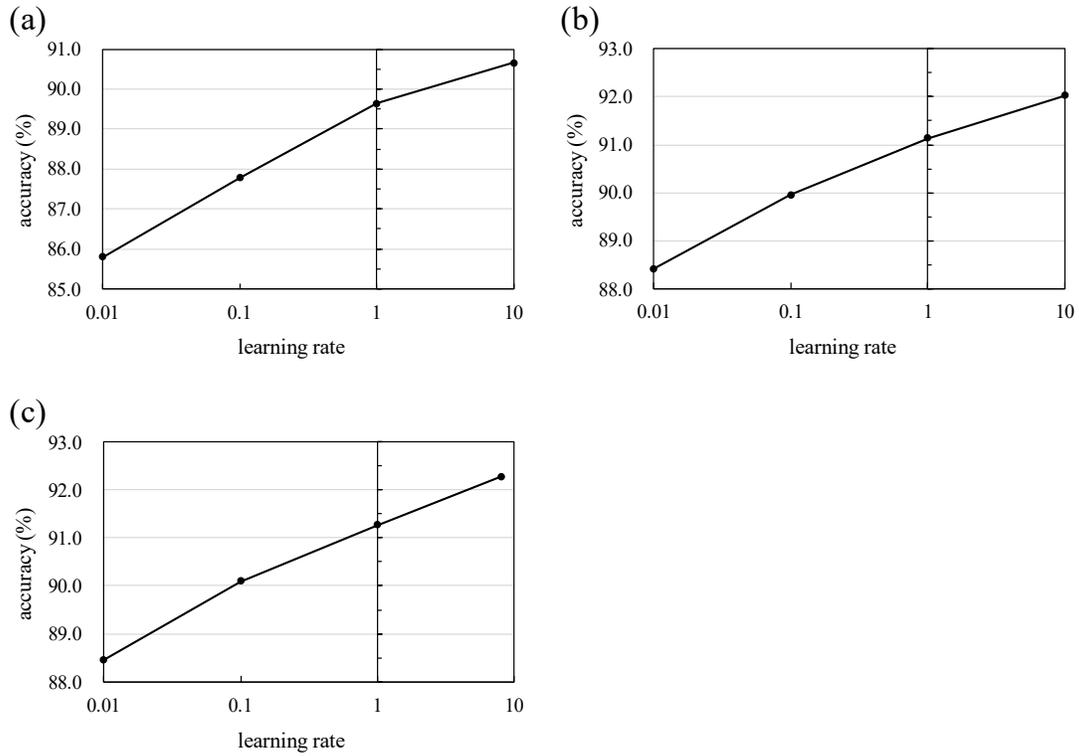

Figure A.4. QCNNs with a nonlinear architecture (12 qubits) using the quantum circuit from Figure 3, evaluated on the FMNIST dataset with (a) 1, (b) 2, and (c) 3 QFilter layers. The x-axis represents the learning rate, while the y-axis denotes accuracy.